\documentclass[journal,comsoc]{IEEEtran}
\usepackage[utf8]{inputenc} 
\usepackage[T1]{fontenc}
\usepackage{url}
\usepackage{ifthen,hyperref}
\usepackage[cmex10]{amsmath}
\usepackage{array} 

\usepackage{enumerate}
\usepackage{amssymb}
\usepackage{amsmath} 

\usepackage{algorithm}
\usepackage{algorithmicx}
\usepackage{algpseudocode} 

\usepackage{hhline}
\usepackage{mathrsfs}
\usepackage{bm}
\usepackage{tikz}
\usetikzlibrary{arrows}
\usepackage[caption=false]{subfig}
\usepackage{graphicx,booktabs,multirow}
\usepackage{epstopdf}
\usepackage{multirow}
\usepackage{tabularx} 
\usepackage{booktabs}
\usepackage{cite}

\usepackage{pifont}

\definecolor{colorhkust}{RGB}{20,43,140}
\definecolor{colortsinghua}{RGB}{116,52,129}
\definecolor{color1}{RGB}{128,0,0}

\usepackage{amsthm}

\theoremstyle{definition}

\theoremstyle{remark}

\begin{document}
      \title{WirelessAgent: Large Language Model Agents for Intelligent Wireless Networks} 
      \author{Jingwen Tong, Wei Guo, Jiawei Shao, Qiong Wu, Zijian Li, Zehong Lin, and Jun Zhang,~\IEEEmembership{Fellow,~IEEE}
      \thanks{This work was supported by the Hong Kong Research Grants Council under the Areas of Excellence scheme grant AoE/E-601/22-R and NSFC/RGC Collaborative Research Scheme grant CRS\_HKUST603/22. 
     The authors are with the Department of Electronic and Computer Engineering, Hong Kong University of Science and Technology, Hong Kong.  (The corresponding author is J. Zhang). }}
        \maketitle
\begin{abstract} 
The rapid evolution of wireless networks presents unprecedented challenges in managing complex and dynamic systems.
Existing methods are increasingly facing fundamental limitations in addressing these challenges. 
In this paper, we introduce \emph{WirelessAgent}, a novel framework that harnesses large language models (LLMs) to create autonomous AI agents for diverse wireless network tasks.  
This framework integrates four core modules that mirror human cognitive processes: perception, memory, planning, and action.
To implement it, we provide a basic usage based on agentic workflows and the LangGraph architecture.
We demonstrate the effectiveness of WirelessAgent through a comprehensive case study on network slicing. 
The numerical results show that WirelessAgent achieves $44.4\%$ higher bandwidth utilization than the \emph{Prompt-based} method, while performing only $4.3\%$ below the \emph{Rule-based optimality}. Notably, WirelessAgent delivers near-optimal network throughput across diverse network scenarios. 
These underscore the framework's potential for intelligent and autonomous resource management in future wireless networks.
The code is available at \url{https://github.com/jwentong/WirelessAgent_R1}.

\end{abstract}

\section{Introduction}
Wireless communications have become a cornerstone of modern society, profoundly impacting daily life and driving innovation across industries. The evolution of network architectures toward hyper-dense and heterogeneous has introduced unprecedented complexity in managing wireless tasks. Contemporary methodologies, such as model-driven optimization and data-driven machine learning techniques, are increasingly facing fundamental limitations in addressing these challenges. These critical bottlenecks necessitate paradigm-shifting innovations in network intelligence frameworks to meet the stringent requirements of future wireless networks \cite{letaief2019roadmap}.

In recent years, enhancing the intelligence of 6G networks has emerged as a consensus for managing growing network complexity and achieving unparalleled performance \cite{bariah2024large}. However, existing AI-based methods in wireless communications are often tailored to specific problems and lack generalizability \cite{tong2024federated}. For example, traditional machine learning models heavily rely on labeled data and task-specific training, making them ill-suited for dynamic network environments or diverse scenarios \cite{shao2024wirelessllm}. This limitation highlights the urgent need for versatile and generalizable AI methods to address the challenges in wireless networks.

The advent of ChatGPT and subsequent advancements in Large Language Models (LLMs) have garnered significant attention due to their ability to understand and generate human-like context, reason across domains, and integrate vast amounts of pre-trained knowledge \cite{brown2020language}. Despite their remarkable capabilities, the direct application of LLMs to wireless networks poses unique challenges. For instance,  standalone LLMs struggle to process multi-modal data (e.g., channel state information (CSI)), dynamically decompose complex tasks, and interact with specialized tools. These limitations hinder the practical deployment of LLMs in complex wireless environments.

To overcome these challenges, LLM-based \emph{AI agents} have emerged as a promising paradigm \cite{xi2025rise}. By augmenting LLMs with modular capabilities, such as environmental data perception and external tool integration, these AI agents enable long-term planning and autonomous decision-making. This integration allows LLMs to function as intelligent entities capable of learning, reasoning, and executing actions in dynamic wireless environments. As a result, LLM-based AI agents can handle diverse wireless tasks with high reliability and efficiency.

In this paper, we introduce \emph{WirelessAgent}, an innovative framework that leverages LLMs to construct autonomous AI agents capable of addressing diverse wireless tasks. It is built on three key supports: LLMs, external tools, and knowledge bases, and operates on three fundamental principles: interaction capability, autonomy, and self-improvement mechanisms. Based on this foundation, the WirelessAgent integrates four core modules: perception, memory, planning, and action, emulating human cognitive processes to effectively manage complex wireless tasks. The perception module processes multi-modal inputs, while the memory module maintains contextual awareness across operations. The planning module facilitates reasoning and decision-making, and the action module executes the proposed solutions. Together, these modules enable continuous feedback loops and dynamic interactions with the external environment.

Although recent studies have explored the application of LLM-based AI agents in wireless networks, a significant gap persists between conceptual designs and practical implementations \cite{shinn2023reflexion,durante2024agent}. To bridge this gap, we provide a basic usage of WirelessAgent based on agentic workflows and the LangGraph architecture. It first determines the agentic workflow for an unfamiliar task in a human-in-the-loop dialogue with LLM application programming interfaces (APIs). Then, the agentic workflow is built on the LangGraph architecture and stored to process routine tasks. 
The resulting agentic workflows achieve a balance between computational efficiency and robust performance, meeting the stringent requirements of high reliability and low latency essential for complex wireless tasks.

We demonstrate the effectiveness of WirelessAgent through a comprehensive case study on network slicing. We conduct extensive simulations on network slicing to evaluate and compare WirelessAgent with the \emph{Prompt-based} and \emph{Rule-based} methods. The Prompt-based method relies solely on prompt engineering for network slicing without the integration of external tools, whereas the Rule-based method assumes optimal performance from an oracle's perspective. Numerical results show that WirelessAgent achieves $44.4\%$ higher bandwidth utilization than the Prompt-based method, while performing only $4.3\%$ below the Rule-based optimality. Furthermore, WirelessAgent supports up to $25$ users, compared to only $15$ users with the Prompt-based approach. Most importantly, WirelessAgent demonstrates near-optimal network throughput across diverse network scenarios. These results highlight that WirelessAgent can intelligently and autonomously manage network slicing tasks while maintaining robustness across varying network conditions.

\subsection{Related Work}
\textbf{LLMs for Wireless Networks:} LLMs have already been specifically investigated for direct use in wireless communication networks to improve their intelligence. Given their inherent strength in natural language processing (NLP), the first applications of LLMs in wireless communications lie in NLP-related communication tasks, such as telecom language understanding and semantic analysis \cite{erak2024leveraging, maatouk2024large,maatouk2023teleqna, dandoush2024large}, where these LLMs excel at processing technical language, identifying user intents, and synthesizing complex information. For instance, the authors in \cite{dandoush2024large} have demonstrated the capability of LLMs to interpret configuration parameters and optimize communication protocols based on contextual understanding. Besides, recent works have shown that pre-trained LLMs can be fine-tuned to adapt other modalities in wireless communications \cite{zhou2024large,liu2024llm4cp}. In particular, the authors in \cite{liu2024llm4cp} fine-tuned pre-trained GPT-2 model \cite{radford2019language} for downlink channel prediction tasks and showed superior performance in terms of prediction accuracy and generalizability. However, the majority of these works are limited to isolated applications, leaving the broader potential of LLMs in practical network operations underexplored. This gap highlights the need for further research into integrating LLMs to address the multi-faceted challenges of real-world wireless networks, such as dynamic resource allocation and multimodal data processing.

\textbf{AI Agents for Wireless Networks:} Existing studies on AI agents in wireless networks have primarily focused on developing task-specific models to address individual problems, such as spectrum management, power control, and network planning  \cite{liang2019spectrum, zappone2019wireless, quan2025large}. However, these agents, relying on either model-driven optimization or data-driven machine learning techniques, often require domain-specific training and labeled datasets, and despite their success in specific scenarios, they lack the generalizability and adaptability needed for complex and evolving wireless environments. Recent advancements in AI agents, particularly those incorporating reinforcement learning and few-shot learning, have shown promise in enabling more autonomous and flexible decision-making \cite{shinn2023reflexion,durante2024agent, mongaillard2024large}. However, these methods are still constrained by their inability to process multimodal data, dynamically decompose complex tasks, and interact seamlessly with domain-specific tools. This underscores the importance of frameworks like WirelessAgent, which aim to leverage the generalization capabilities of LLMs while integrating modular functionalities for planning, reasoning, and action.

\subsection{Contributions and Paper Structure}
Our main contributions are summarized as follows:
\begin{itemize}
    \item We put forth \textit{WirelessAgent}, a framework that empowers LLMs with four core modules: perception, memory, planning, and action, capable of managing diverse wireless tasks.
    \item We introduce a basic usage to implement WirelessAgent based on agentic workflows and the LangGraph architecture. The agentic workflows meet the stringent requirements of high reliability and low latency in handling complex wireless tasks.    
    \item We provide a proof-of-concept case study for the network slicing task, which demonstrates the effectiveness of WirelessAgent in accurately understanding user intent, effectively allocating slice resources, and consistently maintaining optimal performance.
\end{itemize}

The remainder of this paper is organized as follows. Section \ref{Sec_AWN} defines three key supports and principles for AI agents in wireless networks. The WirelessAgent framework and basic usage are proposed in Sections \ref{Sec_WF} and \ref{Sec_BU}, respectively. Section \ref{Sec_CS} presents the WirelessAgent-enabled network slicing as a case study. The numerical results are shown in Section \ref{Sec_NR}. Finally, the paper concludes in Section \ref{Sec_CD}.

\section{Agents for Wireless Networks}\label{Sec_AWN}

\begin{figure*}[!t]
    \centering    \includegraphics[width=1\textwidth]{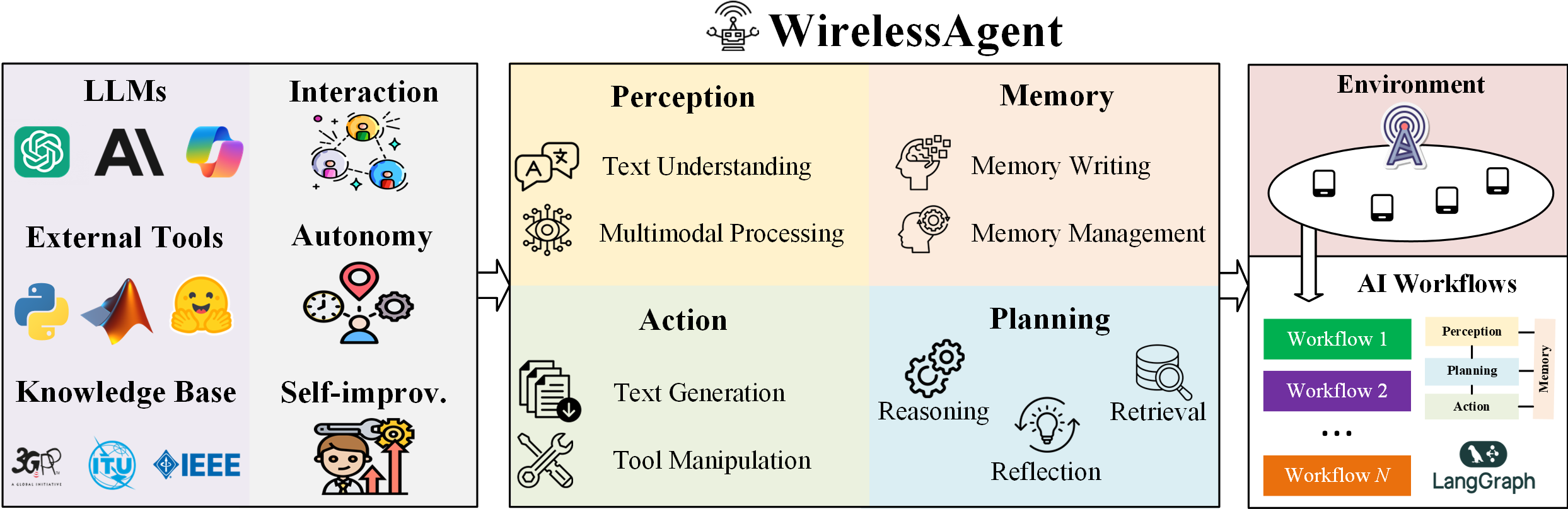}
    \caption{The overview of the WirelessAgent. From left to right, the three parts are foundation, core modules, and basic usage. The foundation, given in Section \ref{Sec_AWN}, includes three key supports (LLMs, external tools, and knowledge base) and three principles (interaction, autonomy, and self-improvement). The core modules of the WirelessAgent framework, introduced in Section \ref{Sec_WF}, consist of perception, memory, action, and planning modules. The basic usage, given in Section \ref{Sec_BU}, provides a practical way to implement the WirelessAgent based on agentic workflows and the LangGraph architecture. } \label{Fig:WirelessAgent}
\end{figure*}

This section provides a foundation for the AI agents for wireless networks. As shown in Fig. \ref{Fig:WirelessAgent}, we first identify three key supports for AI agents, including LLMs, external tools, and the knowledge base. Then, we define three principles for AI agents in handling wireless tasks, consisting of interaction, autonomy, and self-improvement. Next, we elaborate on these key supports and principles.

The WirelessAgent is built upon three essential key supports. The LLMs serve as the cognitive engine, providing sophisticated language understanding, reasoning capabilities, and contextual processing that allow the agent to interpret complex instructions and generate appropriate responses. The external tools extend the agent's functional capabilities by incorporating specialized wireless solutions such as beamforming algorithms and signal processing methods. The knowledge base complements these components by supplying domain-specific information. These three key supports create a powerful foundation that enables WirelessAgent to perceive multimodal inputs, process complex wireless scenarios, and autonomously execute sophisticated network management tasks with high precision and reliability.

\subsection{Supports of Agents for Wireless Networks}

\textbf{LLMs.} LLMs have become indispensable for understanding and generating human-like text from provided inputs. This capability is foundational for developing AI agents that can interpret and respond to complex instructions from humans and other agents \cite{xi2023rise}. LLMs serve as the cognitive core for AI agents, enabling advanced functionalities through techniques such as multimodal perception and tool utilization, which significantly broaden their operational scope across various domains. The ability of LLMs to perform few-shot and zero-shot generalization further enhances their adaptability, allowing them to tackle new tasks without extensive retraining. Furthermore, fine-tuning LLMs with domain-specific knowledge, such as communication standards, patents, and publications, can substantially improve their performance in handling complex wireless tasks.
These promising capabilities underscore the utility of AI agents in wireless systems, facilitating the development of intelligent and autonomous networks \cite{xu2024large}. 

\textbf{External tools.} 
External tools are a critical component in enhancing the capabilities of AI agents, extending their intrinsic functionalities to address specialized wireless tasks. These tools include external datasets, web searches, and APIs that provide additional data and operational capabilities beyond the native abilities of LLMs. However, the complexity of wireless communication tasks necessitates the integration of domain-specific tools that go beyond general-purpose solutions. To address these demands, AI agents incorporate specialized tools such as ray-tracing simulations, beamforming algorithms, and advanced signal processing methods. These tools ensure the accurate and efficient execution of wireless tasks. By leveraging such specialized resources, external tools not only enhance the agent’s ability to handle complex tasks but also significantly strengthen its autonomous decision-making capabilities, providing robust technical support for each operational sub-task.

\textbf{Knowledge base.} 
The knowledge base serves as the cornerstone of AI agents, acting as a comprehensive repository of information. General knowledge bases for agents include diverse resources that provide essential contextual and factual data to support task execution. However, AI agents designed for wireless networks utilize a specialized wireless knowledge base that encompasses industry standards, communication protocols, and cutting-edge research papers. This domain-specific repository enriches the agent’s understanding of complex wireless scenarios. By leveraging this specialized knowledge base, agents can make more accurate decisions in dynamic and challenging wireless environments.

\subsection{Principles of Agents for Wireless Networks}

\textbf{Interaction.} 
An essential principle of AI agents for wireless networks is their ability to effectively interact with humans, the environment, and other agents. These interactions must be adaptive, allowing the agent to understand and respond to various communication modulations and protocols. Furthermore, agents in wireless networks should seamlessly interface with diverse wireless systems to collect information, control parameters, and optimize performance based on the specific requirements of each scenario. Collaboration among agents is another critical aspect of interaction, as it facilitates improved decision-making and resource management. By sharing sensory data, computational resources, and network bandwidth, agents can execute tasks more efficiently in heterogeneous and dynamic network environments.

\textbf{Autonomy.} 
Autonomy is a defining characteristic of AI agents, enabling them to operate independently without continuous human intervention. Effective agents must have significant control over their actions and internal states. While they should be capable of following explicit human instructions, agents should also proactively develop strategies and complete tasks without requiring detailed step-by-step guidance. In the context of wireless networks, agents must respond rapidly to immediate changes or failures in telecommunication systems. By autonomously detecting and addressing issues, agents can implement effective countermeasures and adapt operational strategies to ensure uninterrupted performance, even in the face of unexpected disruptions.

\textbf{Self-improvement.} 
The ability to continually learn and adapt is a crucial requirement for AI agents in the rapidly evolving field of wireless communications. Agents should incorporate mechanisms to learn from interactions, feedback, and changes in the environment, enabling them to remain effective over time. This involves dynamically updating their knowledge base with new data and refining their capabilities to address emerging challenges. Advanced AI techniques, such as reinforcement learning and prompt engineering, play a pivotal role in enabling self-improvement. By evolving over time, agents enhance their intelligence, utility, and ability to maintain optimal performance in increasingly complex wireless scenarios.

\section{WirelessAgent Framework}\label{Sec_WF}
Building on these key supports and principles, we present \emph{WirelessAgent}, a framework leveraging LLMs to develop AI agents capable of managing complex wireless tasks. 
As shown in Fig. \ref{Fig:WirelessAgent}, the WirelessAgent integrates four core modules that mirror human cognitive processes: perception, memory, planning, and action.
Next, we elaborate on these four modules.

\subsection{Perception}
The perception module is adept at processing and understanding diverse forms of input, mirroring the capabilities of human sensory organs. It includes two key functionalities:

\subsubsection{Text understanding}
Language serves as a rich medium for communication, encapsulating extensive information.
Leveraging the advanced capabilities of LLMs, agents can proficiently engage in multi-language understanding and exhibit in-depth comprehension abilities.
During interactions between users and WirelessAgent, textual instructions are provided to LLM agents, including explicit requests and implied intentions.
In addition, by fine-tuning language models with specific datasets and professional corpora, WirelessAgent can interpret complex terminologies used in the domain of wireless communications, bridging the gap between technical language and user-friendly explanations.

\subsubsection{Multimodal processing}
WirelessAgent can autonomously perceive the surrounding environment using equipped sensors and collect multimodal data, encompassing 2D/3D vision and radio signals \cite{yin2023survey}.
Although LLMs exhibit outstanding performance in language conversations, they cannot inherently analyze multimodal data. 
The multimodality contains a wealth of information, including properties of objects, spatial relationships, and wireless channel conditions.
Such rich information offers the agent a broader context and a more precise understanding, deepening the perception of the environment.
To help LLMs process multimodal data, a straightforward approach is to generate corresponding text descriptions.
This approach is highly interpretable and does not require an additional training phase.

\subsection{Memory}
The memory mechanism empowers WirelessAgent to comprehensively analyze past and current data, enhancing its ability to manage dynamic information in wireless intelligence applications.
After the observations are perceived, a part of them will be stored by the agent for further usage through the memory writing operation.
Besides, the past mistakes, successful interventions, and learned behaviors derived from these experiences are recorded for future reference.
For wireless applications, when a user reports a connectivity issue in a specific location, WirelessAgent stores details like the user's identity (ID), CSI, network conditions, and the troubleshooting steps. 
This information is then structured and indexed for future access.
In addition, organizing and indexing historical records helps in efficient data retrieval and reduces the memory footprint.

\subsection{Planning}
Planning is a key management function that facilitates complex task completion by organizing thoughts, outlining steps, and monitoring progress.
Typically, it involves three key modules: a reasoning module, a retrieval module, and a reflection module. 

\subsubsection{Reasoning module}
Reasoning is crucial in human intellectual activities such as problem solving, decision making, and critical analysis. 
Similarly, for LLM agents, reasoning is essential for addressing complex tasks. 
They should break down complex tasks into manageable sub-tasks and formulate corresponding strategies.
The representative techniques that empower WirelessAgent to perform reasoning include in-context learning (ICL) and chain-of-thought (CoT) prompting.
ICL leverages the demonstrations provided within a prompt and analyzes the information presented in the immediate context to generate responses. 
Given the current state and parameters of a network, such as signal strength, noise levels, and user mobility, ICL can predict potential points of failure or recommend adjustments to optimize throughput and stability.
Besides, CoT reasoning explicitly prompts LLMs to generate intermediate steps or reasoning paths. 
In dynamic spectrum quality-of-service (QoS) requirements, the available resources, and interference levels, WirelessAgent can outline a logical pathway that leads to optimal spectrum utilization strategies.

\subsubsection{Retrieval module}
Retrieving the most appropriate content from either its internal memory or an external knowledge base is crucial for WirelessAgent to enhance the response quality.
Specifically, retrieval-augmented generation (RAG) \cite{RAG_lewis2020retrieval} gives LLMs access to information beyond their training data, retrieving extra sources to ground language models on the most relevant and up-to-date information.
The antecedent experience stored in the internal memory serves as a rich source of learned behaviors and previously encountered scenarios.
This internal knowledge allows the model to quickly adapt to similar situations in future interactions.
Besides, external knowledge integration is essential in fast-evolving fields and dynamic environments.
When queried about the latest 5G standards, the agent will prioritize retrieving the most recent information from 3GPP technical specifications and white papers over older documents.
It will also consider the relevance of each document to the specific aspect of 5G being inquired about, such as the physical layer, network architecture, or security features.

\subsubsection{Reflection module}
A reflection module is incorporated into WirelessAgent to emulate the cognitive learning process inherent in human decision making.
The reflection is divided into reflection-before-action and reflection-after-action manners, each serving distinct purposes to enhance the decisions. 
The reflection-before-action manner is thinking through a situation before making decisions or taking actions.
It involves reflecting on the relationship between the observations and the resultant wireless network performance, drawing connections between the provided information and the outcomes. 
The reflection-after-action manner analyzes the outcomes after an incident or an action. 
It examines past decisions, tracking actions and the subsequent results to learn from past successes or mistakes.
By purposefully designing prompts, WirelessAgent reflects on past network management decisions.

\subsection{Action}
In constructing WirelessAgent, the action module carries out specific commands to interact with the environment.

\subsubsection{Text generation}
The advancement of Transformer-based generative models has equipped LLM agents with language generation capabilities.
However, these language models tend to hallucinate or generate text that is fluent and plausible but factually incorrect.
This hallucination problem can be especially concerning when applying LLMs to domains like wireless communications, where consistency with the underlying physical principles and system constraints is critical.
There are many approaches to mitigate the hallucination issue and make WirelessAgent follow instructions better.
For instance, fine-tuning LLMs on wireless datasets such as technical publications, patents, and standards can significantly reduce the frequency of hallucinations. 
Besides, alignment techniques like reinforcement learning from human feedback can be employed to further train the models based on specific user interactions and corrections.

\subsubsection{Tool manipulation}
Tools enhance the capabilities of their users. 
When confronted with complex tasks, humans use tools to simplify the process and boost efficiency, which saves time and resources. While LLMs have extensive knowledge from the training data, they can sometimes misinterpret ambiguous prompts or even generate hallucinations.
Specialized tools, such as Python, Matlab, and Huggingface platforms, help LLMs improve their performance, adapt to specific domains, and meet the unique requirements of those domains in a modular way \cite{shen2024hugginggpt}. 
For example, propagation prediction software, which uses the computationally efficient ray-tracing algorithm in Matlab software, generates accurate and explainable simulations of electromagnetic wave behavior in various environments.
This is crucial in fields like telecommunications, where understanding signal propagation can directly impact the design and optimization of networks.

\section{Basic Usage}\label{Sec_BU}

This section provides a basic usage of WirelessAgent based on agentic workflows and the LangGraph architecture. WirelessAgent cannot work alone without clearly defined roles, objectives, and working modes. agentic workflows fill this gap by integrating the key supports and principles, and formalizing the sequence of tasks and decision points to coordinate the agent's activities and drive them toward their intended goals. These workflows enable WirelessAgent to develop precise and well-structured strategies for handling complex or unfamiliar tasks by leveraging its reasoning, retrieval, and reflection capabilities. 

\begin{figure}[!t]
    \centering
    \includegraphics[width=0.48\textwidth]{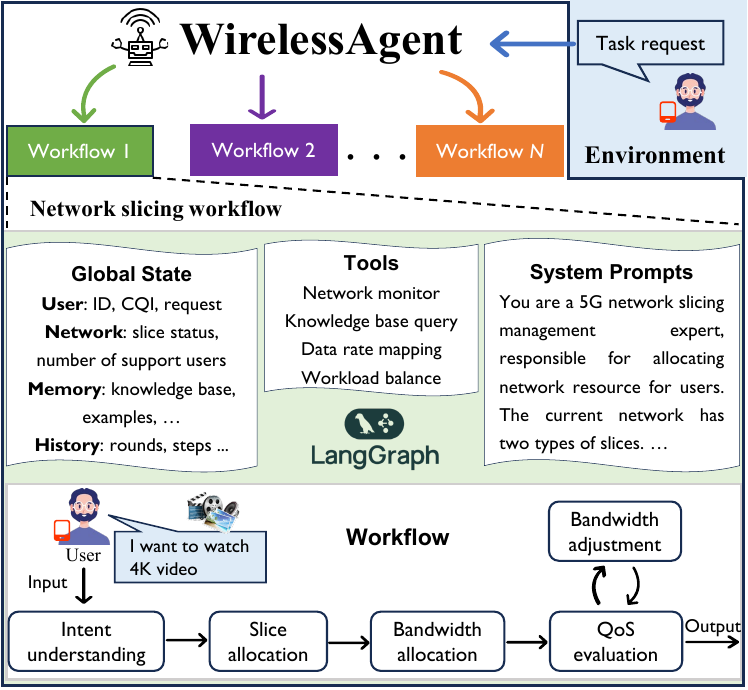}
    \caption{An overview of the basic usage of WirelessAgent.}
    \label{Fig:BU}
\end{figure}

Fig.~\ref{Fig:BU} provides an overview of the basic usage of WirelessAgent, illustrating how the framework processes and manages tasks based on agentic workflows. WirelessAgent receives task requests from the environment and routes them through appropriate workflows. We see that multiple workflows (e.g., Workflow 1, 2, etc.) are deployed to address different tasks, with the network slicing workflow highlighted as an example. Each agentic workflow maintains key components, including a global state, external tools, system prompts, and workflow, all of which are streamlined under the LangGraph architecture. Next, we introduce the workflow determination process and the streamlined LangGraph architecture.

\subsection{Workflow Determination Process}

The workflow determination process relies on advanced reasoning and reflection functions to decompose new or unfamiliar tasks into stable and professional workflows. This process is achieved through repeated interactions with advanced LLMs (e.g., DeepSeek-R1) and human experts in a human-in-the-loop dialogue. 

The process begins with the systematic decomposition of complex wireless tasks into several sub-tasks, identifying task dependencies, and required external tools. The effectiveness of this task decomposition process relies on a dual cognitive mechanism between reflection and action. The \textit{reflection-before-action} mechanism ensures that workflows are evaluated for constraint satisfaction, while the \textit{reflection-after-action} mechanism reviews the outcomes of previous workflows to incorporate learned insights. This dual cognitive approach is further enhanced by repeated prompt engineering, ICL, and expert feedback. In this way, we can identify the objectives, decision logic, and required external tools for each sub-task. This iterative loop ensures that the resulting workflows meet user requirements, satisfy QoS constraints, and are semantically precise for real-world wireless environments.

\subsection{LangGraph Architecture}

LangGraph is a graph-based architectural framework integrating LLMs to map, organize, and execute complex workflows \cite{langchain2024langgraph}. It represents workflows as a network of interconnected nodes, each corresponding to a specific function or sub-task, and edges capture logical and semantic relationships between them. The required global state is central to LangGraph's operation, and it serves as a shared memory repository that persists across the entire workflow execution. This global state maintains critical information, including workflow progress, intermediate results, configuration parameters, and contextual data that must be accessible to all nodes. Based on the global state, external tools, and system prompts, LangGraph enables seamless information sharing between components while preserving the execution history needed for coherent decision-making throughout the workflow lifecycle.

This architecture dynamically adapts workflows by incorporating real-time performance metrics and feedback, ensuring that the allocated resources meet predefined QoS standards. LangGraph facilitates the efficient management of multi-step processes, making it a versatile tool for optimizing wireless tasks. For instance, as illustrated in Figure~\ref{Fig:BU}, LangGraph streamlines the network slicing workflow by interpreting user intents, recommending suitable slices, allocating necessary resources, and maintaining system stability through workload balancing. This modular and adaptable framework ensures the seamless execution of complex and routine operations.

In summary, this basic usage of WirelessAgent operates through agentic workflows and LangGraph architecture, requiring structured workflows to formalize task sequences. Its workflow determination process leverages reasoning and reflection capabilities to decompose complex wireless tasks into manageable sub-tasks, identifying appropriate tools through interactions with advanced LLMs and expert feedback. LangGraph serves as the architectural foundation, organizing workflows as interconnected nodes and maintaining global states to make decisions at each sub-task. This approach enables WirelessAgent to handle complex and routine operations precisely. Next, we demonstrate its effectiveness in network slicing tasks by autonomously allocating slices and bandwidth to satisfy diverse QoS demands.

\section{Case Study: network slicing}\label{Sec_CS}

This section provides a proof-of-concept case study on the network slicing task to demonstrate the effectiveness of the WirelessAgent. 
Network slicing is a technology in 5G that allows creating multiple virtual networks on a shared physical infrastructure. 
Each virtual network is customized to meet specific QoS requirements for different applications \cite{li2018deep}. 
There are typically three types of network slices: Enhanced Mobile Broadband (eMBB) focuses on high data rates and capacity; Ultra-Reliable Low-Latency Communications (URLLC) provides extremely low latency and high reliability, ideal for applications like autonomous vehicles; Massive Machine-Type Communications (mMTC) supports many low-power, low-data-rate IoT devices. 

Without loss of generality, we consider two types of slices (i.e., URLLC and eMBB), where the available physical resource blocks (PRBs) for each slice are represented by bandwidth.
The objective of the network slicing task is to autonomously allocate slices and bandwidth to users sequentially based on their requests and CSI. 
The WirelessAgent-based network slicing begins by identifying a robust agentic workflow for the network slicing task. Once the workflow is determined, the generated workflow code, built on the LangGraph architecture, is executed to complete this task.

\begin{figure}[!t]
    \centering
    \includegraphics[width=0.48\textwidth]{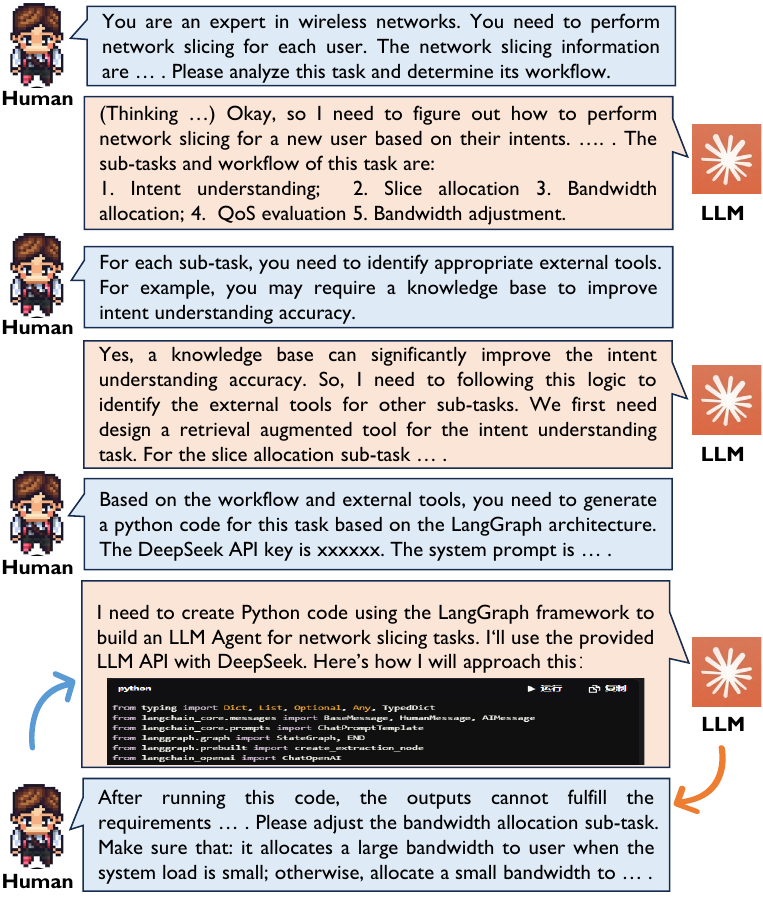}
    \caption{The process of network slicing workflow determination in the human-in-the-loop dialogue.}
    \label{Fig:SLM_NS}
\end{figure}
\subsection{Network Slicing Workflow}
Next, we show how to determine the network slicing workflow through iterative interactions with an expert and an advanced LLM under the human-in-the-loop dialogue. 
As shown in Fig.~\ref{Fig:SLM_NS}, the network slicing task is decomposed into five sub-tasks after several iterations: intent understanding, slice allocation, bandwidth allocation, QoS evaluation, and bandwidth adjustment. Thereafter, it identifies the external tools and logic required for each sub-task. Based on this agentic workflow and external tools, it generates executable code for the network slicing task built on the LangGraph architecture. By repeatedly interacting with the advanced model, the workflow code can be enforced to solve the task more accurately and completely.

We illustrate the role and function of each sub-task and the logic between them. Initially, a user is labeled with its ID, channel quality indicator (CQI), and request. When a user arrives, the WirelessAgent will do:
\begin{enumerate}
    \item \textit{Intent understanding:} This sub-task analyzes the user's request. To improve accuracy, a knowledge base is utilized, containing network slicing examples collected from various sources. This assists the LLM in understanding the user's request.
    \item \textit{Slice allocation:} Based on the analysis results, the agent recommends a suitable slice type (e.g., eMBB or URLLC). This step leverages the LLM's capability to interpret the context and make informed decisions.
    \item \textit{Bandwidth allocation:} This sub-task determines the user’s transmission rate by allocating appropriate bandwidth based on the user’s CQI and the network's current state. The LLM must understand the resource bounds and transmission rate requirements for the specific slice. An external tool, such as a CQI-MCS (modulation and coding scheme) mapping function, maps the CQI to the transmission rate. From an optimization perspective, this sub-task attempts to solve the following problem:
    \begin{subequations}\small \label{SysGol}
        \begin{align}
        & \underset{}{\max\limits_{\mathbf{B}}}
        & &  \sum_{i=1}^{N_e} \Gamma_{\mathrm{embb}}(B_i)  + \sum_{j=1}^{N_u} \Gamma_{\mathrm{urllc}}(B_j)   \qquad  \label{SG_1}\\
        & \mathrm{s.t.}
        & & \Gamma(B_n) = \alpha B_n \log_{10}\left(1 + 10^{\frac{\eta_n}{10}}\right),  \label{SG_2}\\
        & & &  \sum_{i=1}^{N_e} B_i + \sum_{j=1}^{N_u} B_j = B, \label{SG_3}\\
        & & &  N_e + N_u = N,\\
        & & &  B_e^{\min}\leq B_i \leq B_e^{\max},\\
        & & &  B_u^{\min} \leq B_j \leq B_u^{\max},\\
        & & &  \Gamma_e^{\min}\leq \Gamma_{\mathrm{embb}}(B_i) \leq \Gamma_e^{\max},\\
        & & & \Gamma_u^{\min}\leq \Gamma_{\mathrm{urllc}}(B_j) \leq \Gamma_u^{\max},\label{SG_8}
        \end{align}
    \end{subequations}
    where $N$ and $B$ denote the total number of users and the total available bandwidth, respectively. $\Gamma_{\mathrm{embb}}(B_i)$ and $\Gamma_{\mathrm{urllc}}(B_j)$ represent the throughput of eMBB user $i$ and URLLC user $j$, respectively, with allocated bandwidths $B_i$ and $B_j$. Constraint~\eqref{SG_2} calculates the data rate of user $n$, where $\alpha$ is an engineering coefficient and $\eta_n$ is the CQI of user $n$. Constraints on bandwidth and data rates are enforced by $B_e^{\min}$, $B_e^{\max}$, $\Gamma_e^{\min}$, $\Gamma_e^{\max}$ for eMBB users, and $B_u^{\min}$, $B_u^{\max}$, $\Gamma_u^{\min}$, $\Gamma_u^{\max}$ for URLLC users. Note that we address Problem \eqref{SysGol} using the AI Agent's capabilities rather than solving it directly with convex optimization methods. 
    \item \textit{QoS evaluation:} This sub-task verifies if the allocation results meet all user requirements and constraints \eqref{SG_3}-\eqref{SG_8}. If not, WirelessAgent adjusts resources and may reassign users to different slices during the slice handover step. For instance, a handover or adjustment is necessary if a slice cannot accommodate a new user with higher data rate requirements.
    \item \textit{Bandwidth adjustment:} This step ensures that each slice receives an appropriate share of the network’s resources based on evaluation results. For example, when the URLLC slice is fully occupied, the bandwidth of existing users may be reduced to include additional users.
\end{enumerate}
By integrating these sub-tasks and utilizing the LangGraph-based workflow, WirelessAgent ensures efficient resource allocation, maintains QoS requirements, and adapts dynamically to wireless networks.

\begin{figure*}[!t]
    \centering
    \includegraphics[width=0.98\textwidth]{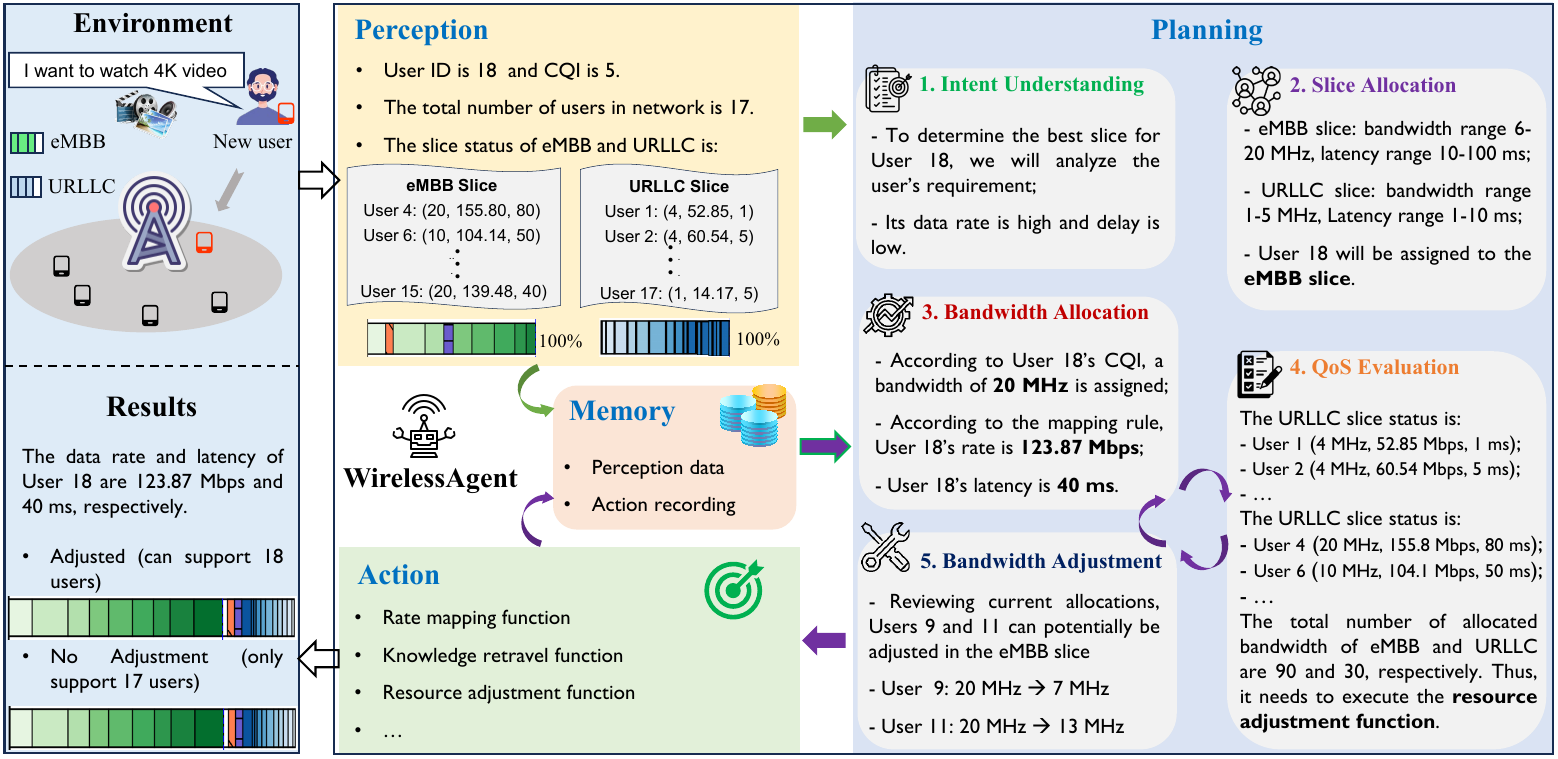}
    \caption{An example of the WirelessAgent-enabled network slicing. It consists of the external and internal environment parts. The external environment includes human interaction and network conditions. The internal environment is the WirelessAgent, where different modules are related to different functions in the network slicing management task. In addition, an example of User 18 is used to visualize the network slicing workflow.} \label{Case_Study}
\end{figure*}

\subsection{Network Slicing Implementation}\label{Sec_BU_NSI}
We implement the WirelessAgent-enabled network slicing in Python using the DeepSeek-V3 API.
We assume that only one user requests service from the BS at each time slot. The total bandwidth is divided into $30$ MHz for the URLLC slice and $90$ MHz for the eMBB slice. The decision ranges for the URLLC and eMBB slices are $[1, 5]$ MHz and $[6, 20]$ MHz, respectively. Each slice type has its own QoS requirements: For the URLLC slice, the transmission rate must be in the range of $[1, 100]$ Mbps, and the latency must satisfy $(0, 10]$ ms. For the eMBB slice, the transmission rate must be in the range of $[100, 400]$ Mbps, and the latency must satisfy $(0, 100]$ ms.

An example of a network slicing task is visualized in Fig.~\ref{Case_Study}, which demonstrates the workflow for managing User 18’s request. The WirelessAgent begins by perceiving information about the user, including ID, location, CQI, and specific requirements. For User 18, whose data rate and latency needs are $123.87$ Mbps and $40$ ms, respectively, the agent assigns the user to the eMBB slice with a bandwidth of $20$ MHz. The memory module stores this allocation data and past actions for future reference. Next, the planning module evaluates current slice status, identifies potential adjustments, and ensures optimal resource allocation. In this case, to support additional users while maintaining QoS, bandwidth adjustments are made for Users 9 and 11 in the eMBB slice. Finally, the action module executes the resource allocation, outputs results, and optimizes the overall network. This example highlights how the agentic workflow ensures efficient and adaptive task execution in dynamic wireless environments.

\section{Numerical Results}\label{Sec_NR}

\begin{figure}[!t]
\centering
\subfloat[Center of HKUST campus]{\label{NetSce_A}
\includegraphics[width=0.44 \columnwidth]{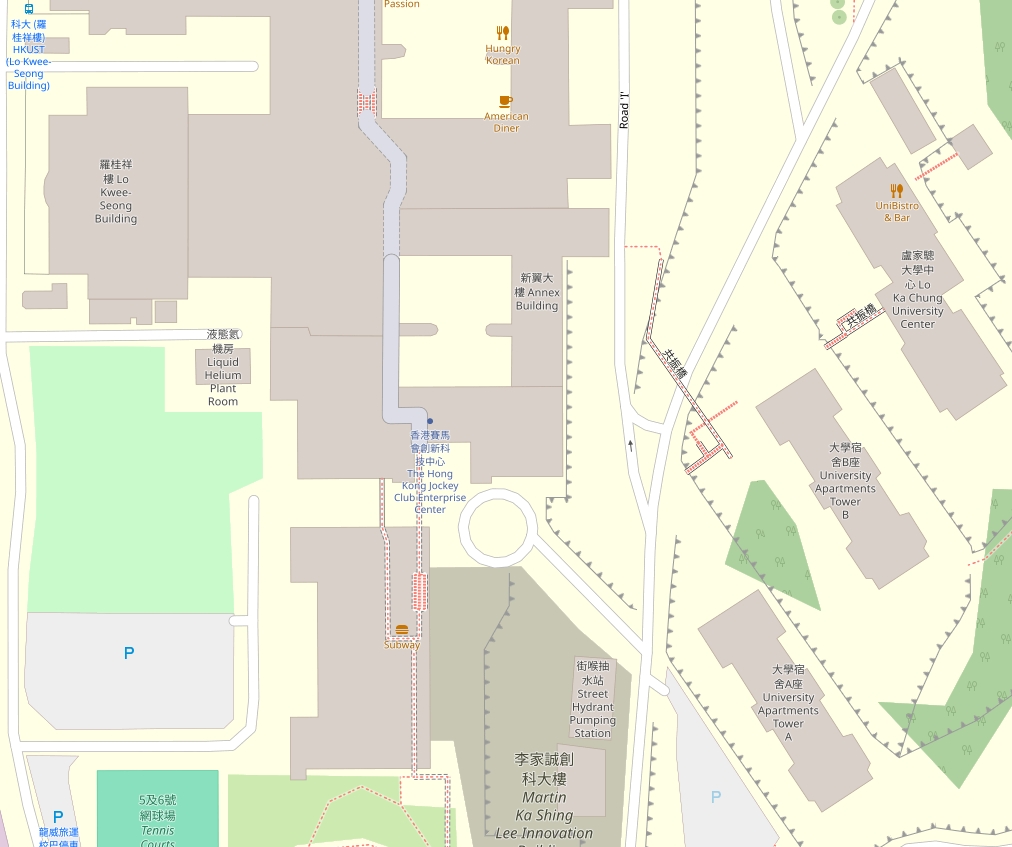}}
\hfil
\subfloat[The received SNRs] {\label{NetSce_B}
\includegraphics[width=0.50 \columnwidth]{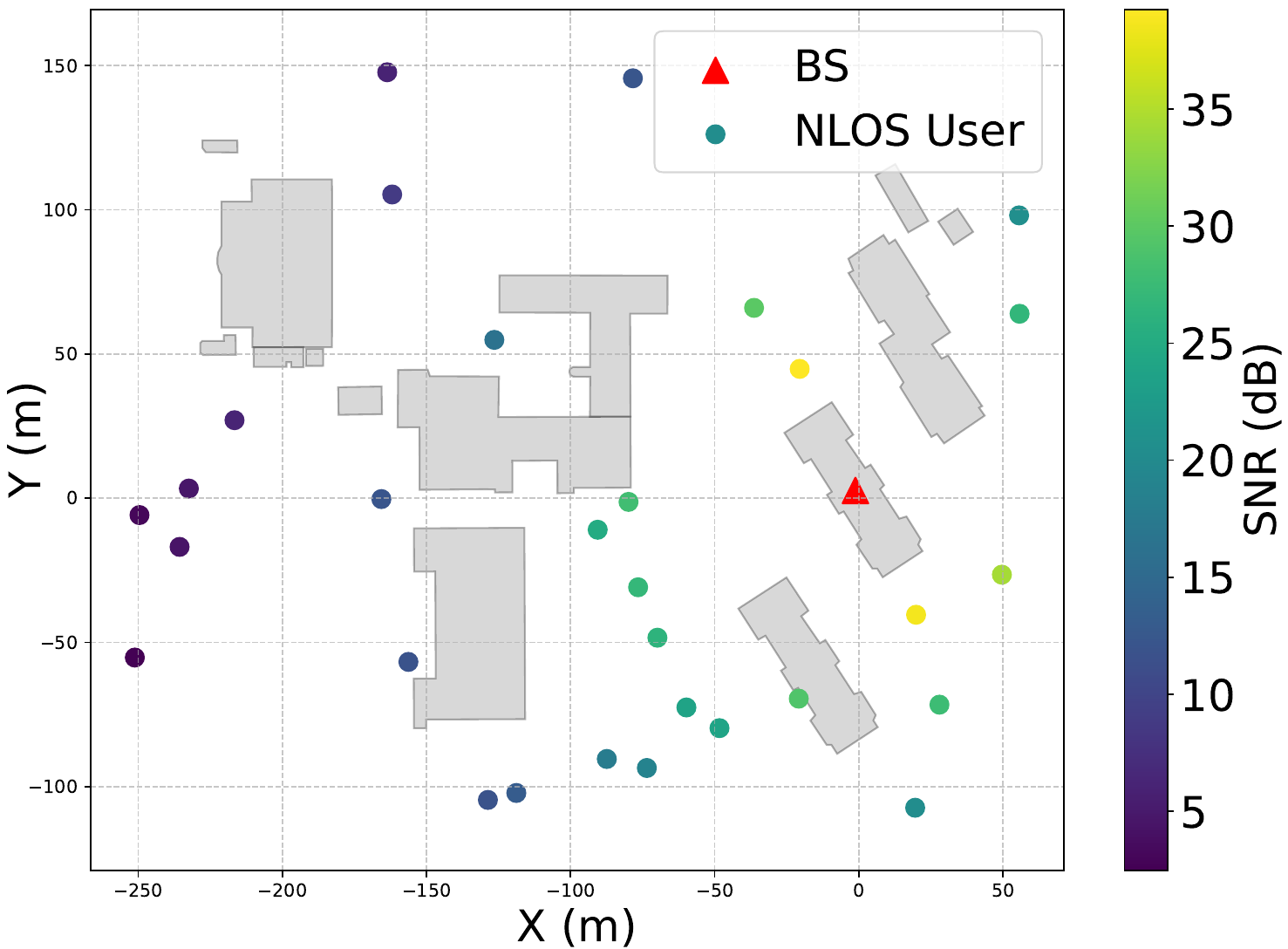}}
\caption{The network scenario. (a) is the center of the HKUST campus screenshot from the OpenStreetMap website; (b) is the received SNRs of the $30$ users on the layout of the left map by performing the ray-tracing algorithm.}
\label{NetSce}
\end{figure}

This section conducts extensive simulations in the network slicing task to evaluate the WirelessAgent under different settings and compare it with two baselines.
We first consider a network scenario located at the center of the HKUST campus, as shown in Fig. \ref{NetSce_A}. There are $N=30$ users uniformly distributed in this area, whose IDs range from $1$ to $30$. The WirelessAgent is deployed at the BS located on the highest building in this area. By performing a ray-tracing algorithm, we obtain the received SNRs and CQIs of the $30$ users, as shown in Fig. \ref{NetSce_B}. The number of ray-tracing paths is set to $6$. Besides, the center frequency of the BS is set to $2.4$ GHz, and the total bandwidth is $B=120$ MHz. The transmission power is $30$ dBm, and the background noise is $-106$ dBm. This ray-tracing step ensures that each user is labeled with an ID and a corresponding CQI. Moreover, users are randomly assigned a request from a dataset collected from the Internet. Other network slicing configurations follow those settings in Section \ref{Sec_BU_NSI}, and all results are obtained from $10$ Monte Carlo trials.

\subsection{Evaluation}

We evaluate the WirelessAgent-enabled network slicing under the aforementioned network scenario. First, we investigate the capabilities of different LLMs within the WirelessAgent. Specifically, we test eight open-source LLMs by calling their APIs, such as DeepSeek-R1, DeepSeek-V3, Llama3.3-70b, Llama3-8b, Qwen-Max-Latest, Qwen-Plus-Latest, Qwen-Turbo-Latest, and QwQ-32b-preview. Second, we compare the intent understanding accuracy of the WirelessAgent when it is equipped with a knowledge base versus when it is not. The intent understanding accuracy is defined as the agreement between the slice recommended by the LLMs and the ground truth.

\begin{table}[t!]
\renewcommand{\arraystretch}{1.3}
\centering
\setlength{\tabcolsep}{4pt}
\begin{tabular}{|l|c|c|c|c|c|}
\hline
\multirow{2}{*}{\textbf{LLMs}} & \multirow{2}{*}{\textbf{Users $\uparrow$}} & \multirow{2}{*}{\textbf{Acc.} $\uparrow$} & \multicolumn{3}{|c|}{\textbf{Average Bandwidth Uti.} $\uparrow$} \\ \cline{4-6} 
          &                       &                       & \textbf{Overall} & \textbf{eMBB} & \textbf{URLLC} \\ \hline
    DeepSeek-R1      &    26      & $100\%$           & $\mathbf{74.65\%}$            & $\mathbf{71.79\%}$    & $82.92\%$           \\ \hline
    DeepSeek-V3      &    26      & $100\%$           & $70.80\%$            & $67.95\%$             &  $79.36\%$           \\ \hline
    Llama3-8b        &    26      & $100\%$           & $60.96\%$            & $62.48\%$             &  $56.41\%$           \\ \hline
    Llama3.3-70b     &    26      & $100\%$           & $74.48\%$            & $\mathbf{71.79\%}$    &  $82.56\%$           \\ \hline
    Qwen-Max    &    26      & $100\%$           & $63.84\%$            & $67.95\%$             &  $51.41\%$           \\ \hline
    Qwen-Plus   &    26      & $100\%$           & $70.74\%$            & $66.75\%$             &  $82.72\%$           \\ \hline
    Qwen-Turbo  &    26      & $100\%$           & $69.36\%$            & $64.96\%$             &  $82.56\%$           \\ \hline
    QwQ-32b     &    26      & $100\%$           & $71.05\%$            & $66.02\%$             &  $\mathbf{86.15\%}$           \\ \hline
\end{tabular}
\caption{The performance of the WirelessAgent-enabled network slicing under different LLMs with a knowledge base.}
\label{tab:WA_KB}
\end{table}

\begin{table}[t!]
\renewcommand{\arraystretch}{1.3}
\centering
\setlength{\tabcolsep}{4pt}
\begin{tabular}{|l|c|c|c|c|c|}
\hline
\multirow{2}{*}{\textbf{LLMs}} & \multirow{2}{*}{\textbf{Users $\uparrow$}} & \multirow{2}{*}{\textbf{Acc.} $\uparrow$} & \multicolumn{3}{|c|}{\textbf{Average Bandwidth Uti.} $\uparrow$} \\ \cline{4-6} 
          &                       &                       & \textbf{Overall} & \textbf{eMBB} & \textbf{URLLC} \\ \hline
    DeepSeek-R1      &    27      & $93.3\%$           & $\mathbf{77.78\%}$            & $\mathbf{76.34\%}$             & $82.10\%$           \\ \hline
    DeepSeek-V3      &    27      & $93.3\%$           & $72.76\%$            & $69.44\%$             & $82.74\%$          \\ \hline
    Llama3-8b        &    27      & $90.0\%$           & $62.40\%$            & $63.87\%$             & $58.02\%$           \\ \hline
    Llama3.3-70b     &    26      & $93.3\%$           & $74.48\%$            & $71.79\%$             & $82.56\%$           \\ \hline
    Qwen-Max    &    26      & $93.3\%$           & $68.56\%$            & $67.05\%$             & $73.08\%$           \\ \hline
    Qwen-Plus   &    27      & $90.0\%$           & $68.22\%$            & $63.92\%$             &  $81.11\%$           \\ \hline
    Qwen-Turbo  &    27      & $93.3\%$           & $70.55\%$            & $66.25\%$             &  $83.46\%$           \\ \hline
    QwQ-32b     &    27      & $90.0\%$           & $77.92\%$            & $73.56\%$             &  $\mathbf{88.51\%}$           \\ \hline
\end{tabular}
\caption{The performance of the WirelessAgent-enabled network slicing under different LLMs without a knowledge base.}
\label{tab:WA_NKB}
\end{table}
Table \ref{tab:WA_KB} presents the supported users, intent understanding accuracy, and average bandwidth utilization of the WirelessAgent under different LLMs with a knowledge base. The term ``supported users'' refers to the maximum number of users that can be supported by the system (or BS). 
The abbreviation ``Acc.'' represents the intent understanding accuracy.
The average bandwidth utilization is defined as the ratio of the bandwidth allocated by the agent to the total available resources, averaged over all time slots. Typically, a higher bandwidth utilization indicates greater system efficiency. We analyze the overall average bandwidth utilization of the system and the average bandwidth utilization of eMBB and URLLC slices separately. 

We observe from Table \ref{tab:WA_KB} that the DeepSeek-R1 model achieves the best performance in terms of overall average bandwidth utilization and the average bandwidth utilization of the eMBB slice among all tested LLMs. The QwQ-32b-preview model achieves the highest average bandwidth utilization in the URLLC slice. In addition, the Qwen-Max-Latest, Qwen-Plus-Latest, and Qwen-Turbo-Latest models exhibit comparable performance. In contrast, the Llama3-8b model suffers from the poorest performance across all metrics. These results demonstrate that the stronger the LLM capability, the better the agent's performance. 

For comparison, Table \ref{tab:WA_NKB} provides similar performance metrics on the WirelessAgent but without a knowledge base. Again, we observe that the deepseek-r1 model achieves the best performance, while the Llama3-8b model suffers from the poorest performance. More importantly, a comparison of Tables \ref{tab:WA_KB} and \ref{tab:WA_NKB} reveals that the intent understanding accuracy of the WirelessAgent is consistently higher when equipped with a knowledge base across all LLMs. This demonstrates that WirelessAgent significantly enhances its intent understanding capability by retrieving relevant information from a specialized database (i.e., the tool manipulation capability).

Interestingly, the average bandwidth utilization and the supported users of the WirelessAgent with a knowledge base are slightly lower than those without a knowledge base for different LLMs. 
These discrepancies arise because some eMBB users are misclassified as URLLC users when the knowledge base is lacking. As a result, the system can support more users since the URLLC users consume less bandwidth than the eMBB users.

\begin{figure*}[!t]
    \centering    \includegraphics[width=0.96\textwidth]{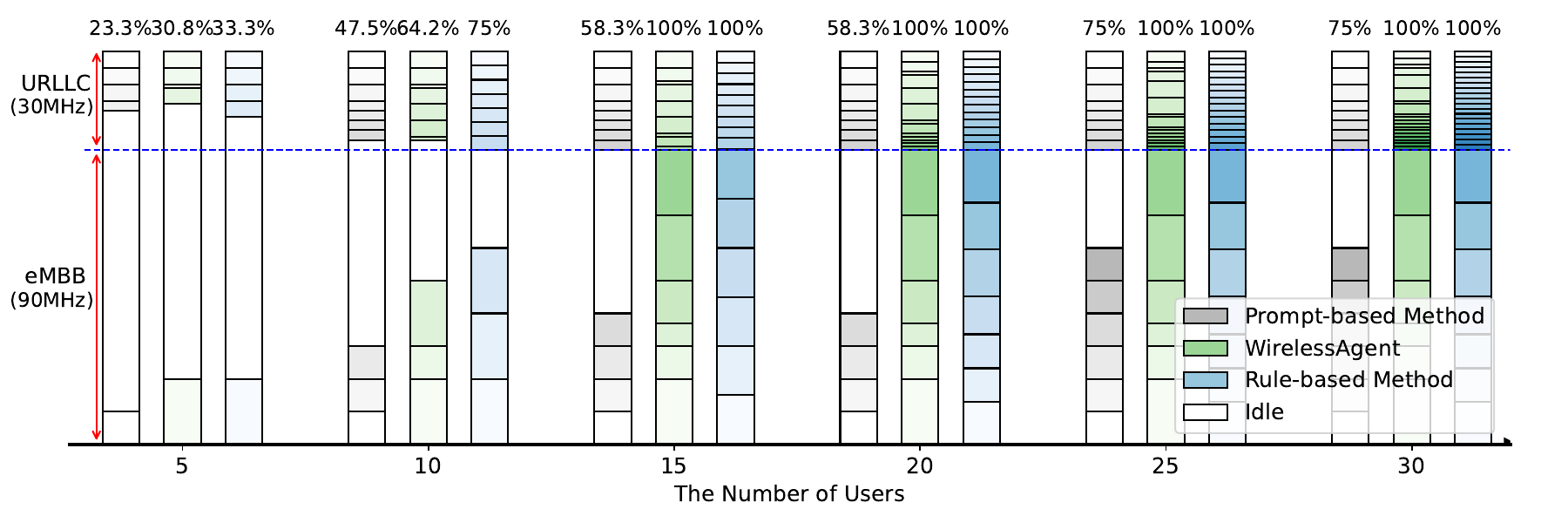}
    \caption{The bandwidth utilization rate of the prompt-based method, WirelessAgent, and the rule-based method for the network slicing task under different numbers of users. The total bandwidth of the eMBB and URLLC slices is $30$ and $90$ MHz, respectively. Each rectangular block in the bar chart represents the bandwidth allocated to one user. For example, there are a total of $6$ users in the eMBB slice for the WirelessAgent when the number of users is $30$.} \label{Fig:Utilization}
\end{figure*}

\begin{figure*}[!t]
\centering
\subfloat[North of HKUST campus]{\label{RayTrace01}
\includegraphics[width=0.65 \columnwidth]{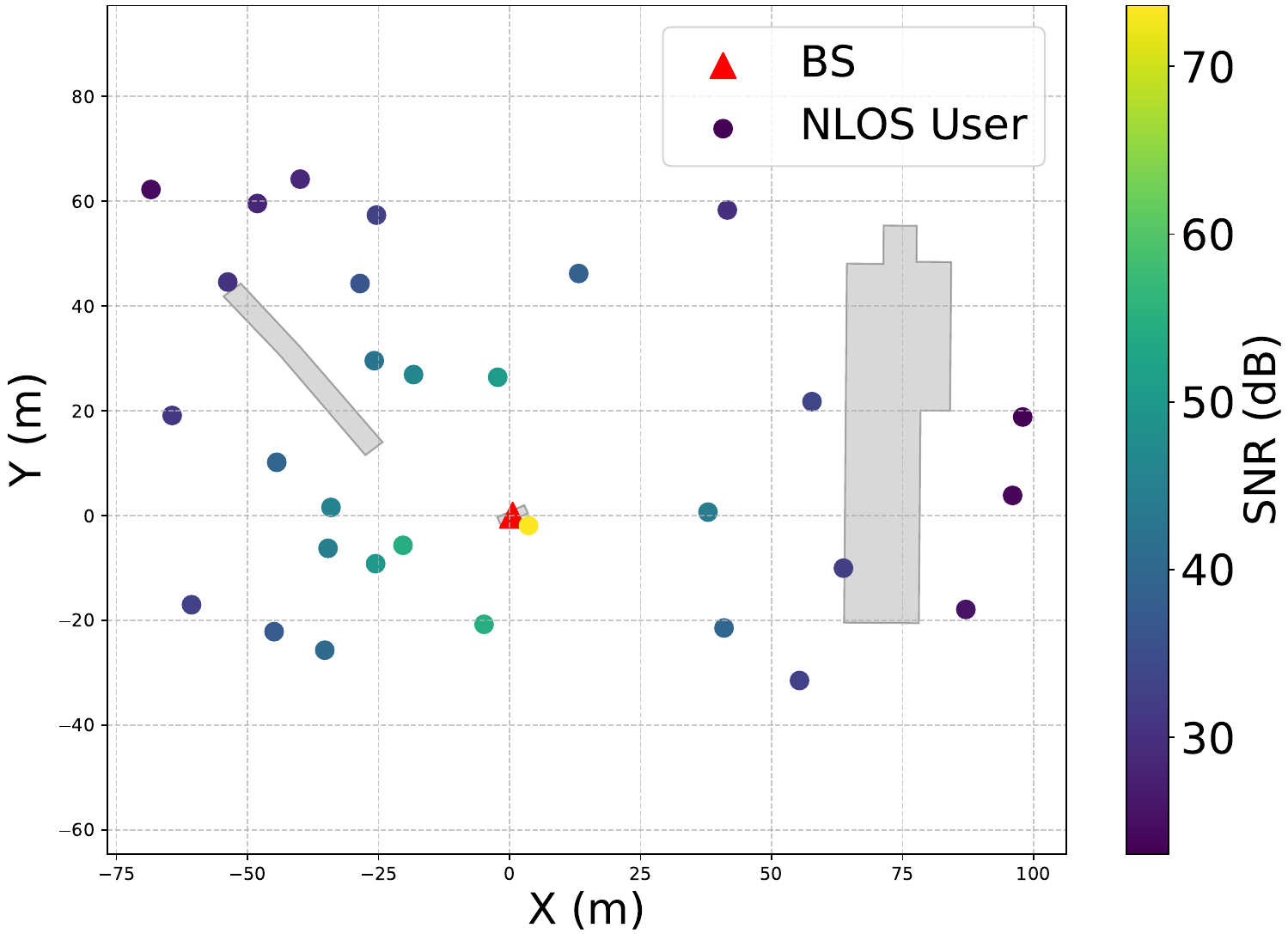}}
\hfil
\subfloat[Center of HKUST campus] {\label{RayTrace02}
\includegraphics[width=0.65 \columnwidth]{Figs/ray_tracing_map_center.pdf}}
\hfil
\subfloat[South of HKUST campus] {\label{RayTrace03}
\includegraphics[width=0.65 \columnwidth]{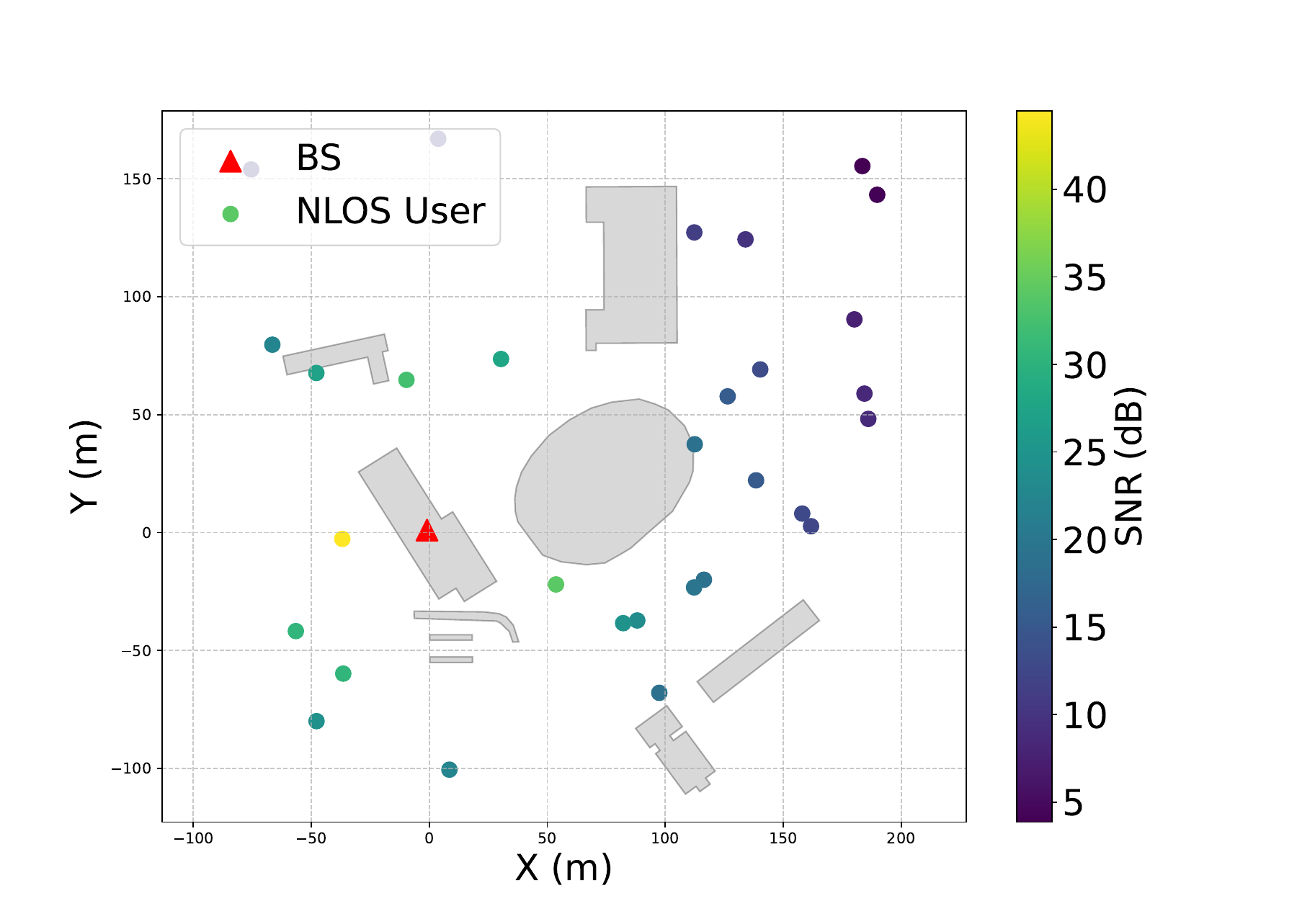}}
\hfil
\subfloat[Performance comparison at scenario (a)] {\label{WirelessAgent01}
\includegraphics[width=0.65 \columnwidth]{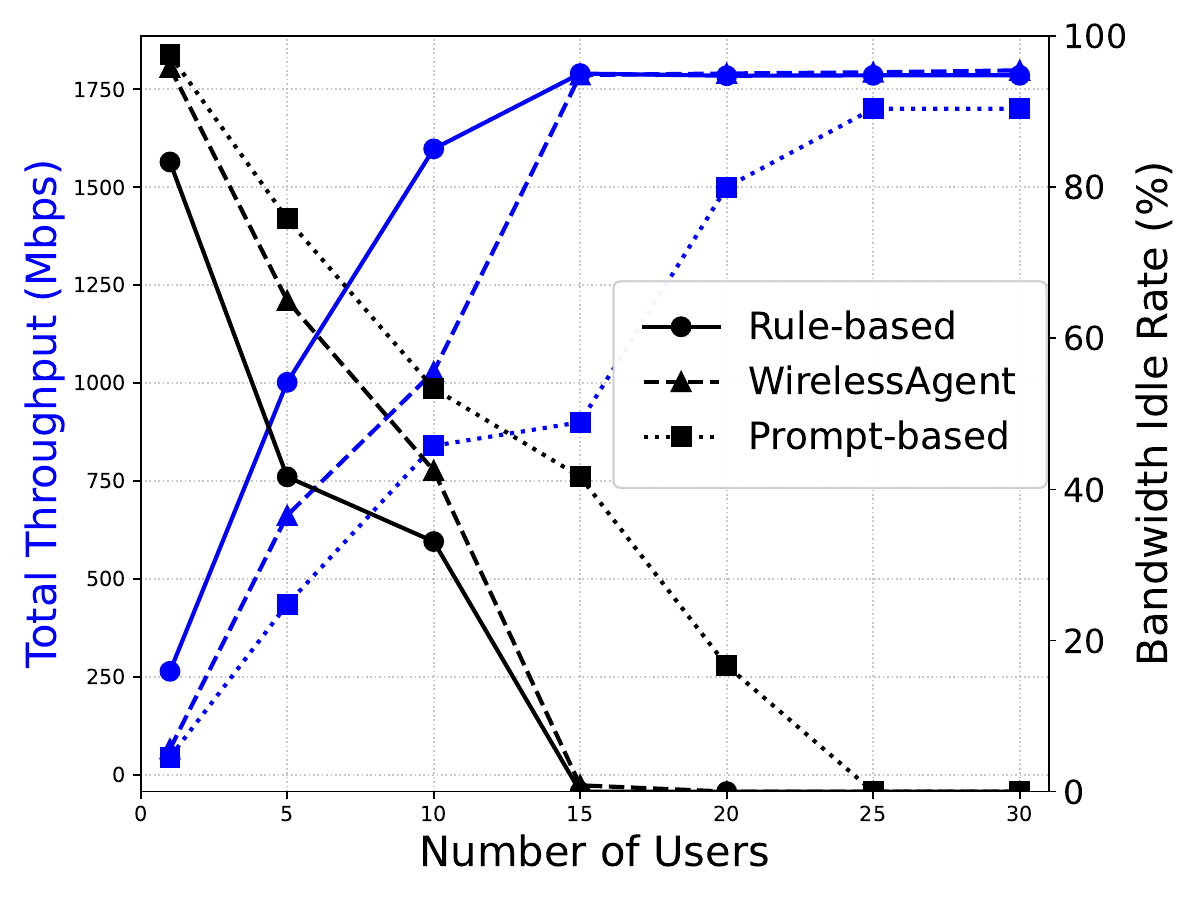}}
\hfil
\subfloat[Performance comparison at scenario (b)] {\label{WirelessAgent02}
\includegraphics[width=0.65 \columnwidth]{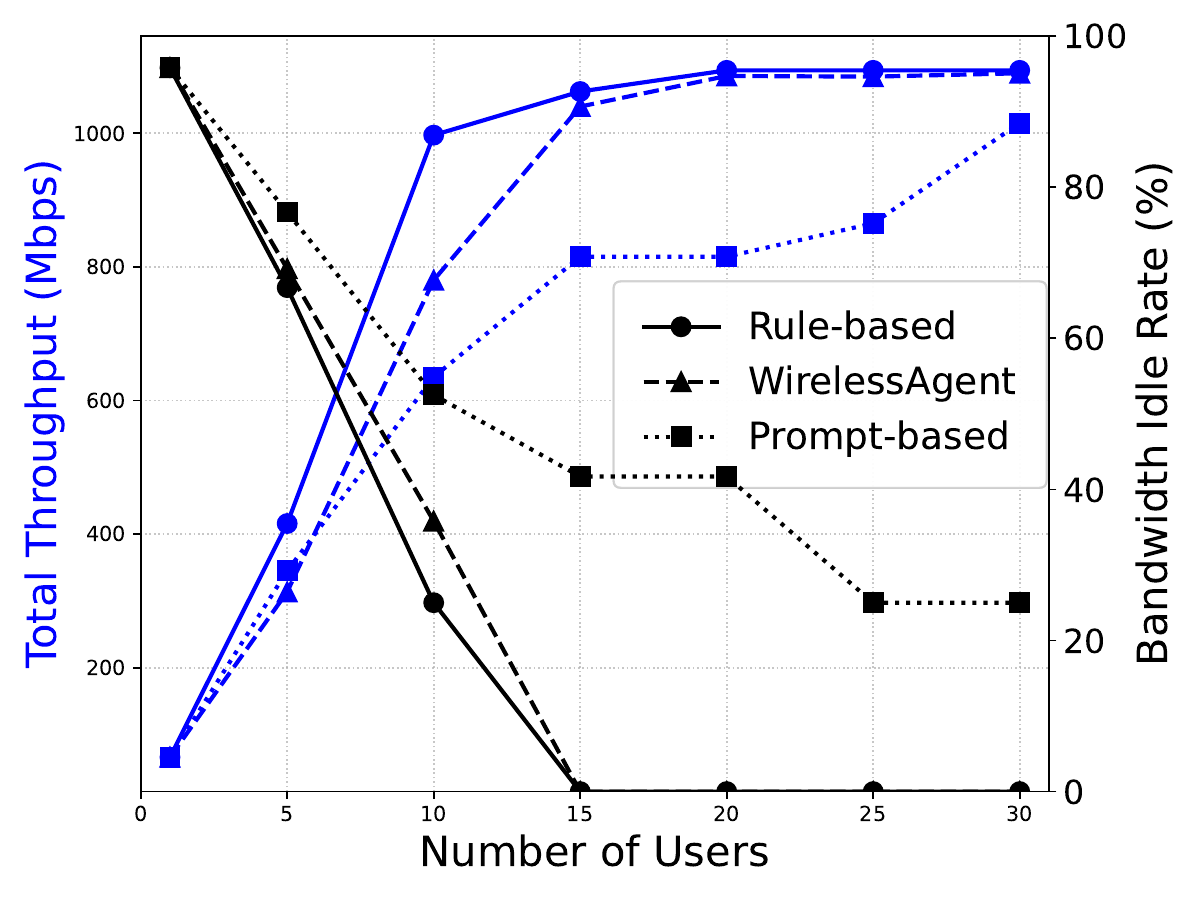}}
\hfil
\subfloat[Performance comparison at scenario (c)] {\label{WirelessAgentM03}
\includegraphics[width=0.65 \columnwidth]{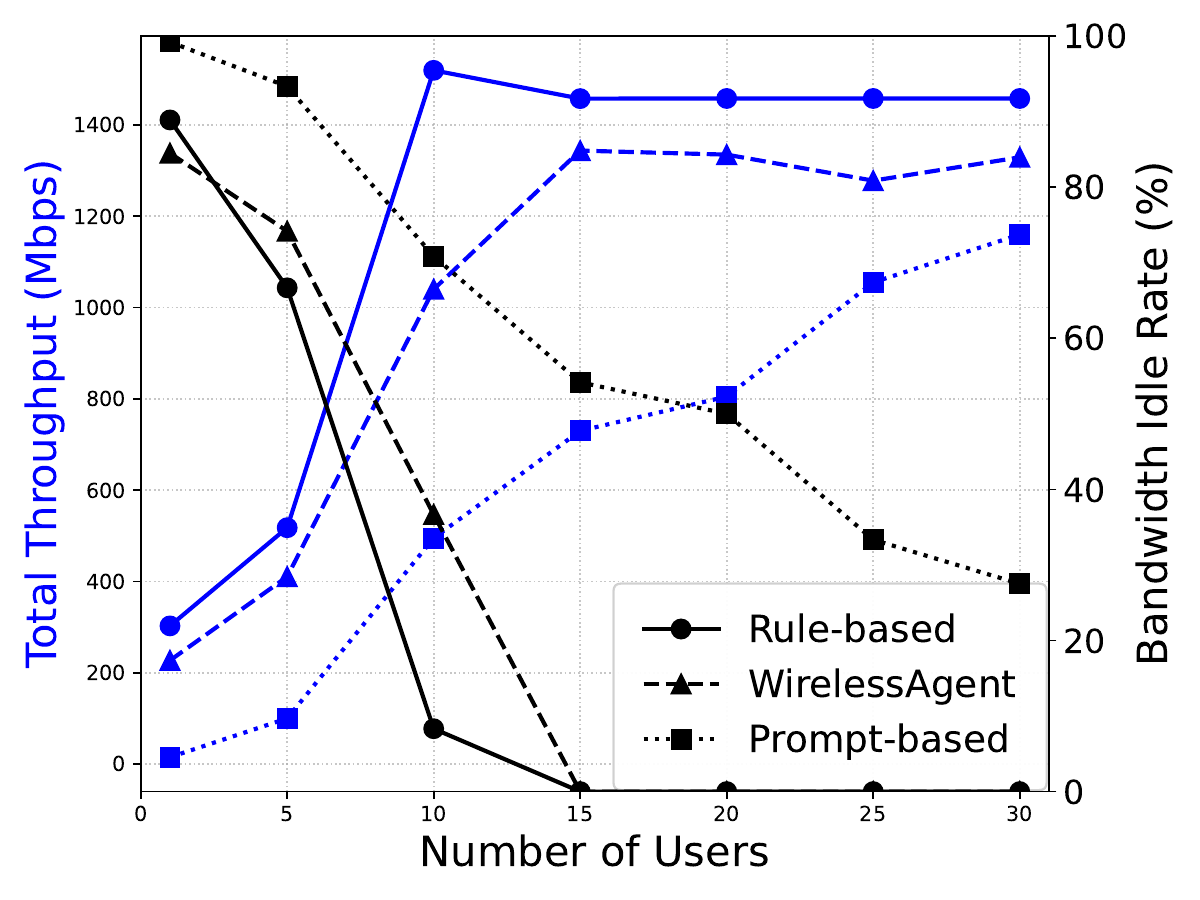}}
\caption{The total throughput and bandwidth idle rate of the Rule-based method, WirelessAgent, and Prompt-based method for the network slicing task via the number of users under three network scenarios. (a)-(c) are the received SNRs of the $30$ users after performing ray tracing under the network scenarios of the north, center, and south of the HKUST campus, respectively. (d)-(f) are the corresponding performance of these methods under the above network scenarios.}
\label{RayTrace_Thr_BIR}
\end{figure*}
\subsection{Comparison}
Next, we compare the \emph{WirelessAgent} with the \emph{Prompt-based} and \emph{Rule-based} methods for the network slicing task under the aforementioned scenario. The WirelessAgent utilizes the DeepSeek-V3 API to execute the workflow. The Prompt-based method refers to an approach that relies solely on prompt engineering for network slicing without integrating any external tools. This method also adopts the DeepSeek-V3 API as its core component for the task. Fig.~\ref{Fig:ImpleDetails} (see Appendix \ref{app:a}) illustrates an example of the Prompt-based network slicing method. We see that the system prompt is carefully designed to guide the task decomposition using the CoT technique. It is worth noting that the WirelessAgent also adopts this system prompt, where its planning process aligns closely with the workflow illustrated in Fig.~\ref{Case_Study}. For the Rule-based method, we assume ideal intent understanding and perfect slice allocation. Then, the network slicing task is formulated as a throughput maximization problem as shown in Problem~\eqref{SysGol}. This problem can be solved efficiently using the convex optimization tools.

Fig.~\ref{Fig:Utilization} shows the bandwidth utilization rates achieved by the Prompt-based method, WirelessAgent, and Rule-based method for the network slicing task under varying numbers of users. As shown in the figure, the Rule-based method achieves the best performance in terms of bandwidth utilization. The WirelessAgent achieves a significantly higher utilization rate than the Prompt-based method and performs very close to the Rule-based method. Additionally, the maximum number of users supported by the Prompt-based, WirelessAgent, and Rule-based methods is $15$, $25$, and $26$, respectively. The inferior performance of the Prompt-based method stems from its frequent violations of the constraints in Problem~\eqref{SysGol}, which occur due to the absence of external tools to enforce these constraints. These observations demonstrate that the WirelessAgent can intelligently and autonomously manage network slicing tasks with near-optimal performance.

Then, we compare the total throughput and bandwidth idle rate achieved by the Rule-based, WirelessAgent, and Prompt-based methods across different numbers of users under three network scenarios: the north, center, and south areas of the HKUST campus. The WirelessAgent uses the DeepSeek-V3 API equipped with external tools and a knowledge base, while the Prompt-based method solely relies on the DeepSeek-V3 API for reasoning and execution. The bandwidth idle rate is defined as the proportion of unused bandwidth to the total available bandwidth. 

From Fig. \ref{RayTrace_Thr_BIR}, we observe that the Rule-based method consistently delivers the best performance across all scenarios, achieving the highest total throughput and the lowest bandwidth idle rate. In contrast, the Prompt-based method exhibits the poorest performance across all metrics and scenarios. Notably, the WirelessAgent consistently outperforms the Prompt-based method in all cases and achieves performance that is very close to the Rule-based method. These results highlight the robustness and adaptability of the WirelessAgent in effectively handling network slicing tasks across diverse scenarios.

\section{Conclusions and Future Works}\label{Sec_CD}
This paper introduced WirelessAgent, a novel framework that leverages LLMs to create autonomous AI agents capable of addressing diverse wireless network tasks. 
The WirelessAgent integrated four cognitive modules (perception, memory, planning, and action) that mirror human cognitive processes. We provided a practical implementation approach based on agentic workflows and the LangGraph architecture, which strikes a balance between computational efficiency and robust performance. Through a comprehensive case study on network slicing, we demonstrated that WirelessAgent achieves $44.4\%$ higher bandwidth utilization than the Prompt-based method while performing only $4.3\%$ below the Rule-based optimality. Furthermore, WirelessAgent supports up to $25$ users compared to only $15$ users with the Prompt-based approach and delivers near-optimal network throughput across diverse network scenarios. These results highlight the significant potential of WirelessAgent as a framework for intelligent, autonomous control in future wireless networks, capable of adapting to dynamic environments while maintaining robust performance. The code is available at \url{https://github.com/jwentong/WirelessAgent_R1}.

Some discussions and promising directions on WirelessAgent are:

\textbf{Agentic Workflow vs. AI Agent:} 
A critical consideration addressed in our study is the choice between a workflow agent and a generic agent. The workflow agent, which is designed to follow a structured sequence of steps or sub-tasks, aligns closely with established network management procedures such as network slicing. It excels in scenarios where tasks need to be decomposed into predictable, repeatable operations, ensuring clear audit trails and easier troubleshooting. In contrast, the generic agent is more reactive and flexible, potentially handling a broader range of unforeseen or dynamic tasks. However, this flexibility can sometimes lead to less predictable outcomes and challenges in reliability, especially under stringent real-time constraints. Our results suggest that the workflow agent model is more suitable for wireless network tasks where adherence to strict QoS requirements and predictable resource allocation is paramount. It provides structured decision-making and better integrates external tools to ensure that each sub-task is performed reliably, thereby delivering robust network performance.

\textbf{Enhanced multimodal integration:} Further research is needed to develop more sophisticated methods for integrating multimodal data into LLMs. This includes exploring advanced encoding techniques and fusion strategies to fully capture the rich information in wireless environments.

\textbf{Explainable wireless AI:} As agents for wireless networks become involved in critical network management decisions, ensuring transparency and explainability in their decision-making process is crucial. Future work should focus on developing methods for interpreting and explaining the reasoning behind agent actions.

\textbf{Security and privacy considerations:} The deployment of agents in real-world wireless networks necessitates robust security and privacy safeguards. Research efforts should address potential vulnerabilities and develop mechanisms to protect sensitive network information and user data.

\textbf{Real-world deployment and evaluation:} Moving beyond simulations, future work should focus on deploying and evaluating agents in real-world wireless networks. This will involve addressing practical challenges related to scalability, robustness, and integration with existing network infrastructure.

\appendix
\begin{figure}[!t]
    \centering    \includegraphics[width=0.48\textwidth]{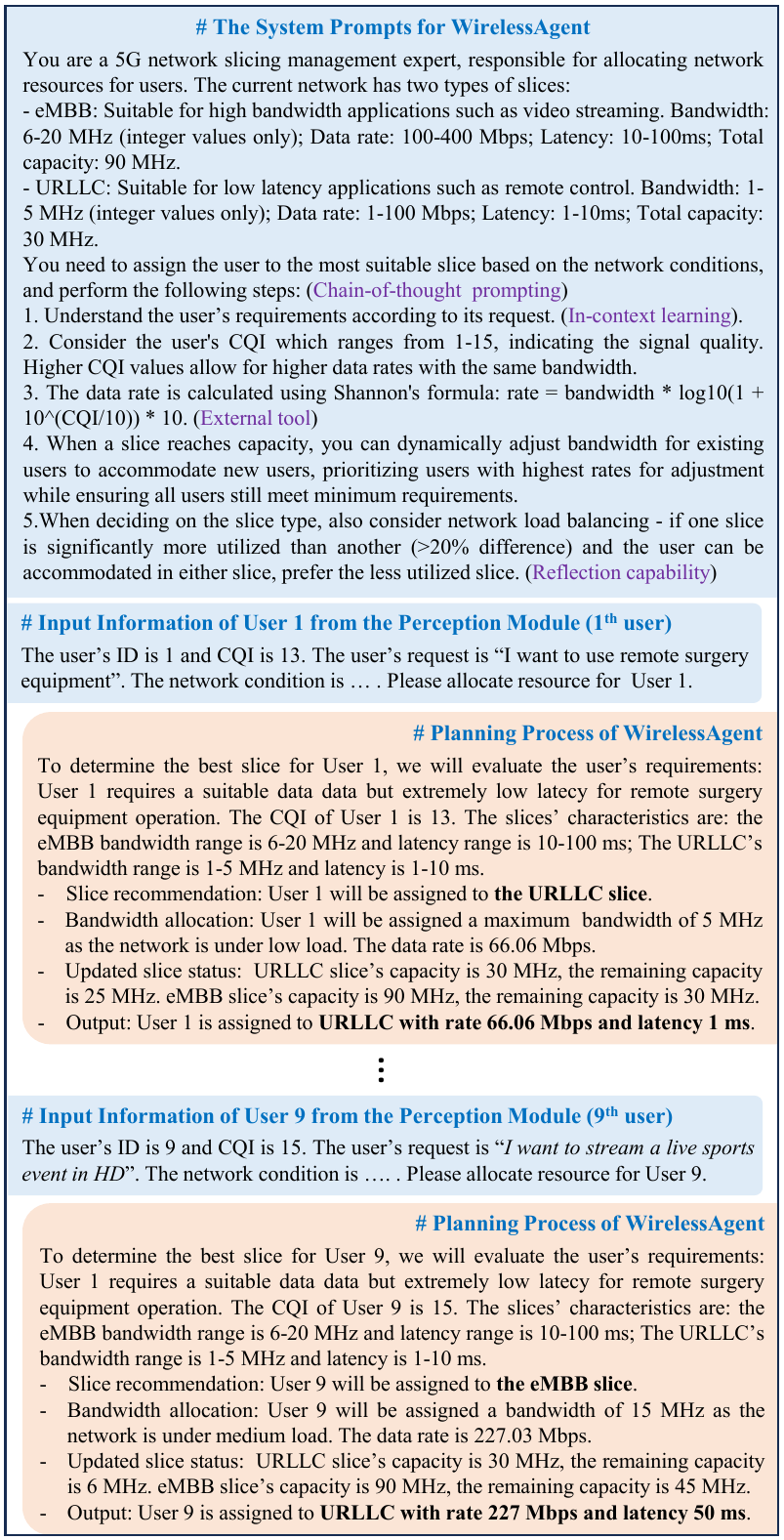}
    \caption{An example of the Prompt-based method used in the network slicing management task.} \label{Fig:ImpleDetails}
\end{figure}
\subsection{An example of the prompt-based method}\label{app:a}
Fig. \ref{Fig:ImpleDetails} shows an example of the prompt-based method for the network slicing task. It can be seen that the users arrive sequentially with information on user ID, CQI, and request. At the beginning, the main prompts are provided to guide the WirelessAgent's workflow, which decomposes this task into multiple sub-tasks using the CoT technique. 
The planning process for two users with different requirements: User 9 requesting HD sports streaming (assigned to eMBB with 15 MHz bandwidth and 227 Mbps data rate) and User 1 needing remote surgery equipment operation (assigned to URLLC with 5 MHz bandwidth and 66.06 Mbps data rate). The system prompt outlines how the agent functions as a network slicing management expert, making decisions based on user requirements, CQI values, and network conditions. The agent intelligently allocates bandwidth within two slice types (eMBB for high bandwidth applications and URLLC for low latency requirements), using Shannon's formula for rate calculations and implementing load balancing techniques to optimize resource utilization while ensuring all users meet their minimum requirements.
It demonstrates that WirelessAgent leverages in-context learning, reflection, and tool manipulation capabilities to complete different sub-tasks. Upon the arrival of a new user, the WirelessAgent initiates the resource allocation task. The planning process aligns closely with the steps outlined in the planning module in Fig. \ref{Case_Study}. This experiment is conducted sequentially until the final user is accommodated.

\bibliographystyle{ieeetr}
\bibliography{myref}

\end{document}